
%
%
%
%
%
\def\standardrisposta{s }\def\reducedrisposta{r }
\def\doublerisposta{d }\def\cartarisposta{e }\def\amsrisposta{y }
\newcount\ingrandimento \newcount\sinnota \newcount\dimnota
\newcount\unoduecol \newdimen\collhsize \newdimen\tothsize
\newdimen\fullhsize \newcount\controllorisposta \sinnota=1
\newskip\infralinea  \global\controllorisposta=0
\message{ ********    Welcome to PANDA macros (Plain TeX, AP, 1991)}
\message{ ******** }
\message{       You'll have to answer a few questions in lowercase.}
\message{>  Do you want it in double-page (d), reduced (r)}
\message{or standard format (s) ? }\read-1 to\risposta
\message{>  Do you want it in USA A4 (u) or EUROPEAN A4 (e)}
\message{paper size ? }\read-1 to\srisposta
\message{>  Do you have AMSFonts 2.0 (math) fonts (y/n) ? }
\read-1 to\arisposta
%
%
%
%
%
\ifx\risposta\standardrisposta \ingrandimento=1200
\message{>> This will come out UNREDUCED << }
\dimnota=2 \unoduecol=1 \global\controllorisposta=1 \fi
\ifx\risposta\reducedrisposta \ingrandimento=1095 \dimnota=1
\unoduecol=1  \global\controllorisposta=1
\message{>> This will come out REDUCED << } \fi
\ifx\risposta\doublerisposta \ingrandimento=1000 \dimnota=2
\unoduecol=2  \global\controllorisposta=1
\message{>> You must print this in LANDSCAPE orientation << } \fi
\ifnum\controllorisposta=0  \ingrandimento=1200
\message{>>> ERROR IN INPUT, I ASSUME STANDARD UNREDUCED FORMAT <<< }
\dimnota=2 \unoduecol=1 \fi
\magnification=\ingrandimento
%
%
%
%
\newdimen\eucolumnsize \newdimen\eudoublehsize \newdimen\eudoublevsize
\newdimen\uscolumnsize \newdimen\usdoublehsize \newdimen\usdoublevsize
\newdimen\eusinglehsize \newdimen\eusinglevsize \newdimen\ussinglehsize
\newskip\standardbaselineskip \newdimen\ussinglevsize
\newskip\reducedbaselineskip \newskip\doublebaselineskip
\eucolumnsize=12.0truecm    
\eudoublehsize=25.5truecm   
\eudoublevsize=6.5truein    
\uscolumnsize=4.4truein     
\usdoublehsize=9.4truein    
\usdoublevsize=6.8truein    
\eusinglehsize=6.5truein    
\eusinglevsize=24truecm     
\ussinglehsize=6.5truein    
\ussinglevsize=8.9truein    
\standardbaselineskip=16pt  
\reducedbaselineskip=14pt   
\doublebaselineskip=12pt    
%
%
\def\Portoffset{}
\def\Landoffset{}
%
%
\def\Landspec{}
\tolerance=10000
\parskip 0pt plus 2pt  
%
%
\ifx\risposta\standardrisposta \infralinea=\standardbaselineskip \fi
\ifx\risposta\reducedrisposta  \infralinea=\reducedbaselineskip \fi
\ifx\risposta\doublerisposta   \infralinea=\doublebaselineskip \fi
\ifnum\controllorisposta=0    \infralinea=\standardbaselineskip \fi
\ifx\risposta\doublerisposta   \Landoffset \else \Portoffset \fi
\ifx\risposta\doublerisposta \ifx\srisposta\cartarisposta
\tothsize=\eudoublehsize \collhsize=\eucolumnsize
\vsize=\eudoublevsize  \else  \tothsize=\usdoublehsize
\collhsize=\uscolumnsize \vsize=\usdoublevsize \fi \else
\ifx\srisposta\cartarisposta \tothsize=\eusinglehsize
\vsize=\eusinglevsize \else  \tothsize=\ussinglehsize
\vsize=\ussinglevsize \fi \collhsize=4.4truein \fi
%
%
%
%
\newcount\contaeuler \newcount\contacyrill \newcount\contaams
\font\ninerm=cmr9  \font\eightrm=cmr8  \font\sixrm=cmr6
\font\ninei=cmmi9  \font\eighti=cmmi8  \font\sixi=cmmi6
\font\ninesy=cmsy9  \font\eightsy=cmsy8  \font\sixsy=cmsy6
\font\ninebf=cmbx9  \font\eightbf=cmbx8  \font\sixbf=cmbx6
\font\ninett=cmtt9  \font\eighttt=cmtt8  \font\nineit=cmti9
\font\eightit=cmti8 \font\ninesl=cmsl9  \font\eightsl=cmsl8
\skewchar\ninei='177 \skewchar\eighti='177 \skewchar\sixi='177
\skewchar\ninesy='60 \skewchar\eightsy='60 \skewchar\sixsy='60
\hyphenchar\ninett=-1 \hyphenchar\eighttt=-1 \hyphenchar\tentt=-1
\def\bfmath{\cmmib}  \font\tencmmib=cmmib10  \newfam\cmmibfam
\skewchar\tencmmib='177  \def\bfcal{\cmbsy} \font\tencmbsy=cmbsy10
\newfam\cmbsyfam  \skewchar\tencmbsy='60
\def\scaps{\cmcsc}  \font\tencmcsc=cmcsc10  \newfam\cmcscfam
\ifnum\ingrandimento=1095 
 
\font\bfone=cmbx10 at 10.95pt

\font\capsone=cmcsc10 at 10.95pt 

\else  
 
\font\bfone=cmbx10 at 12pt

\font\capsone=cmcsc10 at 12pt 
\fi
\def\chapterfont#1{\xdef\ttaarr{#1}}
\def\sectionfont#1{\xdef\ppaarr{#1}}
\def\ttaarr{\bf} \def\ppaarr{\sl}
%
%
\newfam\eufmfam \newfam\msamfam \newfam\msbmfam \newfam\eufbfam
\def\loadamsmath{\global\contaams=1 \ifx\arisposta\amsrisposta
\font\tenmsam=msam10 \font\ninemsam=msam9 \font\eightmsam=msam8
\font\sevenmsam=msam7 \font\sixmsam=msam6 \font\fivemsam=msam5
\font\tenmsbm=msbm10 \font\ninemsbm=msbm9 \font\eightmsbm=msbm8
\font\sevenmsbm=msbm7 \font\sixmsbm=msbm6 \font\fivemsbm=msbm5
\else \def\msbm{\bf} \fi \def\Bbb{\msbm} \def\symbl{\msam} \tenpoint}
\ifx\arisposta\amsrisposta
\font\sevenex=cmex7               
\font\eightex=cmex8  \font\nineex=cmex9
\font\ninecmmib=cmmib9   \font\eightcmmib=cmmib8
\font\sevencmmib=cmmib7 \font\sixcmmib=cmmib6
\font\fivecmmib=cmmib5   \skewchar\ninecmmib='177
\skewchar\eightcmmib='177  \skewchar\sevencmmib='177
\skewchar\sixcmmib='177   \skewchar\fivecmmib='177
\font\ninecmbsy=cmbsy9    \font\eightcmbsy=cmbsy8
\font\sevencmbsy=cmbsy7  \font\sixcmbsy=cmbsy6
\font\fivecmbsy=cmbsy5   \skewchar\ninecmbsy='60
\skewchar\eightcmbsy='60  \skewchar\sevencmbsy='60
\skewchar\sixcmbsy='60    \skewchar\fivecmbsy='60
\font\ninecmcsc=cmcsc9    \font\eightcmcsc=cmcsc8     \else
\def\cmmib{\fam\cmmibfam\tencmmib}\textfont\cmmibfam=\tencmmib
\scriptfont\cmmibfam=\tencmmib \scriptscriptfont\cmmibfam=\tencmmib
\def\cmbsy{\fam\cmbsyfam\tencmbsy} \textfont\cmbsyfam=\tencmbsy
\scriptfont\cmbsyfam=\tencmbsy \scriptscriptfont\cmbsyfam=\tencmbsy
\scriptfont\cmcscfam=\tencmcsc \scriptscriptfont\cmcscfam=\tencmcsc
\def\cmcsc{\fam\cmcscfam\tencmcsc} \textfont\cmcscfam=\tencmcsc \fi
\catcode`@=11
\newskip\ttglue
\gdef\tenpoint{\def\rm{\fam0\tenrm}
  \textfont0=\tenrm \scriptfont0=\sevenrm \scriptscriptfont0=\fiverm
  \textfont1=\teni \scriptfont1=\seveni \scriptscriptfont1=\fivei
  \textfont2=\tensy \scriptfont2=\sevensy \scriptscriptfont2=\fivesy
  \textfont3=\tenex \scriptfont3=\tenex \scriptscriptfont3=\tenex
  \def\mcal{\fam2 \tensy}  \def\mmit{\fam1 \teni}
  \textfont\itfam=\tenit \def\it{\fam\itfam\tenit}
  \textfont\slfam=\tensl \def\sl{\fam\slfam\tensl}
  \textfont\ttfam=\tentt \scriptfont\ttfam=\eighttt
  \scriptscriptfont\ttfam=\eighttt  \def\tt{\fam\ttfam\tentt}
  \textfont\bffam=\tenbf \scriptfont\bffam=\sevenbf
  \scriptscriptfont\bffam=\fivebf \def\bf{\fam\bffam\tenbf}
     \ifx\arisposta\amsrisposta    \ifnum\contaams=1
  \textfont\msamfam=\tenmsam \scriptfont\msamfam=\sevenmsam
  \scriptscriptfont\msamfam=\fivemsam \def\msam{\fam\msamfam\tenmsam}
  \textfont\msbmfam=\tenmsbm \scriptfont\msbmfam=\sevenmsbm
  \scriptscriptfont\msbmfam=\fivemsbm \def\msbm{\fam\msbmfam\tenmsbm}
     \fi  \textfont3=\tenex \scriptfont3=\sevenex
  \scriptscriptfont3=\sevenex
  \def\cmmib{\fam\cmmibfam\tencmmib} \scriptfont\cmmibfam=\sevencmmib
  \textfont\cmmibfam=\tencmmib  \scriptscriptfont\cmmibfam=\fivecmmib
  \def\cmbsy{\fam\cmbsyfam\tencmbsy} \scriptfont\cmbsyfam=\sevencmbsy
  \textfont\cmbsyfam=\tencmbsy  \scriptscriptfont\cmbsyfam=\fivecmbsy
  \def\cmcsc{\fam\cmcscfam\tencmcsc} \scriptfont\cmcscfam=\eightcmcsc
  \textfont\cmcscfam=\tencmcsc \scriptscriptfont\cmcscfam=\eightcmcsc
     \fi            \tt \ttglue=.5em plus.25em minus.15em
  \normalbaselineskip=12pt
  \setbox\strutbox=\hbox{\vrule height8.5pt depth3.5pt width0pt}
  \let\sc=\eightrm \let\big=\tenbig   \normalbaselines
  \baselineskip=\infralinea  \rm}
\gdef\ninepoint{\def\rm{\fam0\ninerm}
  \textfont0=\ninerm \scriptfont0=\sixrm \scriptscriptfont0=\fiverm
  \textfont1=\ninei \scriptfont1=\sixi \scriptscriptfont1=\fivei
  \textfont2=\ninesy \scriptfont2=\sixsy \scriptscriptfont2=\fivesy
  \textfont3=\tenex \scriptfont3=\tenex \scriptscriptfont3=\tenex
  \def\mcal{\fam2 \ninesy}  \def\mmit{\fam1 \ninei}
  \textfont\itfam=\nineit \def\it{\fam\itfam\nineit}
  \textfont\slfam=\ninesl \def\sl{\fam\slfam\ninesl}
  \textfont\ttfam=\ninett \scriptfont\ttfam=\eighttt
  \scriptscriptfont\ttfam=\eighttt \def\tt{\fam\ttfam\ninett}
  \textfont\bffam=\ninebf \scriptfont\bffam=\sixbf
  \scriptscriptfont\bffam=\fivebf \def\bf{\fam\bffam\ninebf}
     \ifx\arisposta\amsrisposta  \ifnum\contaams=1
  \textfont\msamfam=\ninemsam \scriptfont\msamfam=\sixmsam
  \scriptscriptfont\msamfam=\fivemsam \def\msam{\fam\msamfam\ninemsam}
  \textfont\msbmfam=\ninemsbm \scriptfont\msbmfam=\sixmsbm
  \scriptscriptfont\msbmfam=\fivemsbm \def\msbm{\fam\msbmfam\ninemsbm}
     \fi  \textfont3=\nineex \scriptfont3=\sevenex
  \scriptscriptfont3=\sevenex
  \def\cmmib{\fam\cmmibfam\ninecmmib}  \textfont\cmmibfam=\ninecmmib
  \scriptfont\cmmibfam=\sixcmmib \scriptscriptfont\cmmibfam=\fivecmmib
  \def\cmbsy{\fam\cmbsyfam\ninecmbsy}  \textfont\cmbsyfam=\ninecmbsy
  \scriptfont\cmbsyfam=\sixcmbsy \scriptscriptfont\cmbsyfam=\fivecmbsy
  \def\cmcsc{\fam\cmcscfam\ninecmcsc} \scriptfont\cmcscfam=\eightcmcsc
  \textfont\cmcscfam=\ninecmcsc \scriptscriptfont\cmcscfam=\eightcmcsc
     \fi            \tt \ttglue=.5em plus.25em minus.15em
  \normalbaselineskip=11pt
  \setbox\strutbox=\hbox{\vrule height8pt depth3pt width0pt}
  \let\sc=\sevenrm \let\big=\ninebig \normalbaselines\rm}
\gdef\eightpoint{\def\rm{\fam0\eightrm}
  \textfont0=\eightrm \scriptfont0=\sixrm \scriptscriptfont0=\fiverm
  \textfont1=\eighti \scriptfont1=\sixi \scriptscriptfont1=\fivei
  \textfont2=\eightsy \scriptfont2=\sixsy \scriptscriptfont2=\fivesy
  \textfont3=\tenex \scriptfont3=\tenex \scriptscriptfont3=\tenex
  \def\mcal{\fam2 \eightsy}  \def\mmit{\fam1 \eighti}
  \textfont\itfam=\eightit \def\it{\fam\itfam\eightit}
  \textfont\slfam=\eightsl \def\sl{\fam\slfam\eightsl}
  \textfont\ttfam=\eighttt \scriptfont\ttfam=\eighttt
  \scriptscriptfont\ttfam=\eighttt \def\tt{\fam\ttfam\eighttt}
  \textfont\bffam=\eightbf \scriptfont\bffam=\sixbf
  \scriptscriptfont\bffam=\fivebf \def\bf{\fam\bffam\eightbf}
     \ifx\arisposta\amsrisposta   \ifnum\contaams=1
  \textfont\msamfam=\eightmsam \scriptfont\msamfam=\sixmsam
  \scriptscriptfont\msamfam=\fivemsam \def\msam{\fam\msamfam\eightmsam}
  \textfont\msbmfam=\eightmsbm \scriptfont\msbmfam=\sixmsbm
  \scriptscriptfont\msbmfam=\fivemsbm \def\msbm{\fam\msbmfam\eightmsbm}
     \fi  \textfont3=\eightex \scriptfont3=\sevenex
  \scriptscriptfont3=\sevenex
  \def\cmmib{\fam\cmmibfam\eightcmmib}  \textfont\cmmibfam=\eightcmmib
  \scriptfont\cmmibfam=\sixcmmib \scriptscriptfont\cmmibfam=\fivecmmib
  \def\cmbsy{\fam\cmbsyfam\eightcmbsy}  \textfont\cmbsyfam=\eightcmbsy
  \scriptfont\cmbsyfam=\sixcmbsy \scriptscriptfont\cmbsyfam=\fivecmbsy
  \def\cmcsc{\fam\cmcscfam\eightcmcsc} \scriptfont\cmcscfam=\eightcmcsc
  \textfont\cmcscfam=\eightcmcsc \scriptscriptfont\cmcscfam=\eightcmcsc
     \fi             \tt \ttglue=.5em plus.25em minus.15em
  \normalbaselineskip=9pt
  \setbox\strutbox=\hbox{\vrule height7pt depth2pt width0pt}
  \let\sc=\sixrm \let\big=\eightbig \normalbaselines\rm}
\gdef\tenbig#1{{\hbox{$\left#1\vbox to8.5pt{}\right.\n@space$}}}
\gdef\ninebig#1{{\hbox{$\textfont0=\tenrm\textfont2=\tensy
   \left#1\vbox to7.25pt{}\right.\n@space$}}}
\gdef\eightbig#1{{\hbox{$\textfont0=\ninerm\textfont2=\ninesy
   \left#1\vbox to6.5pt{}\right.\n@space$}}}
\def\alternativefont#1#2{\ifx\arisposta\amsrisposta \relax \else
\xdef#1{#2} \fi}
\global\contaeuler=0 \global\contacyrill=0 \global\contaams=0
%
%
%
%
\newbox\fotlinebb \newbox\hedlinebb \newbox\leftcolumn
\gdef\makeheadline{\vbox to 0pt{\vskip-22.5pt
     \fullline{\vbox to8.5pt{}\the\headline}\vss}\nointerlineskip}
\gdef\makehedlinebb{\vbox to 0pt{\vskip-22.5pt
     \fullline{\vbox to8.5pt{}\copy\hedlinebb\hfil
     \line{\hfill\the\headline\hfill}}\vss} \nointerlineskip}
\gdef\makefootline{\baselineskip=24pt \fullline{\the\footline}}
\gdef\makefotlinebb{\baselineskip=24pt
    \fullline{\copy\fotlinebb\hfil\line{\hfill\the\footline\hfill}}}
\gdef\doubleformat{\shipout\vbox{\Landspec\makehedlinebb
     \fullline{\box\leftcolumn\hfil\columnbox}\makefotlinebb}
     \advancepageno}
\gdef\columnbox{\leftline{\pagebody}}
\gdef\line#1{\hbox to\hsize{\hskip\leftskip#1\hskip\rightskip}}
\gdef\fullline#1{\hbox to\fullhsize{\hskip\leftskip{#1}%
\hskip\rightskip}}
\gdef\footnote#1{\let\@sf=\empty
         \ifhmode\edef\#sf{\spacefactor=\the\spacefactor}\/\fi
         #1\@sf\vfootnote{#1}}
\gdef\vfootnote#1{\insert\footins\bgroup
         \ifnum\dimnota=1  \eightpoint\fi
         \ifnum\dimnota=2  \ninepoint\fi
         \ifnum\dimnota=0  \tenpoint\fi
         \interlinepenalty=\interfootnotelinepenalty
         \splittopskip=\ht\strutbox
         \splitmaxdepth=\dp\strutbox \floatingpenalty=20000
         \leftskip=\oldssposta \rightskip=\olddsposta
         \spaceskip=0pt \xspaceskip=0pt
         \ifnum\sinnota=0   \textindent{#1}\fi
         \ifnum\sinnota=1   \item{#1}\fi
         \footstrut\futurelet\next\fo@t}
\gdef\fo@t{\ifcat\bgroup\noexpand\next \let\next\f@@t
             \else\let\next\f@t\fi \next}
\gdef\f@@t{\bgroup\aftergroup\@foot\let\next}
\gdef\f@t#1{#1\@foot} \gdef\@foot{\strut\egroup}
\gdef\footstrut{\vbox to\splittopskip{}}
\skip\footins=\bigskipamount
\count\footins=1000  \dimen\footins=8in
\catcode`@=12
\tenpoint
\ifnum\unoduecol=1 \hsize=\tothsize   \fullhsize=\tothsize \fi
\ifnum\unoduecol=2 \hsize=\collhsize  \fullhsize=\tothsize \fi
\global\let\lrcol=L
\ifnum\unoduecol=1 \output{\plainoutput{\ifnum\tipbnota=2
\clearnmbnota\fi}} \fi
\ifnum\unoduecol=2 \output{\if L\lrcol
     \global\setbox\leftcolumn=\columnbox
     \global\setbox\fotlinebb=\line{\hfill\the\footline\hfill}
     \global\setbox\hedlinebb=\line{\hfill\the\headline\hfill}
     \advancepageno  \global\let\lrcol=R
     \else  \doubleformat \global\let\lrcol=L \fi
     \ifnum\outputpenalty>-20000 \else\dosupereject\fi
     \ifnum\tipbnota=2\clearnmbnota\fi }\fi
\def\ifdoublepage{\ifnum\unoduecol=2 }
\gdef\yespagenumbers{\footline={\hss\tenrm\folio\hss}}
\gdef\ciao{\par\vfill\supereject \ifnum\unoduecol=2
     \if R\lrcol  \headline={}\nopagenumbers\null\vfill\eject
     \fi\fi \end}
\newskip\olddsposta \newskip\oldssposta
\global\oldssposta=\leftskip \global\olddsposta=\rightskip
\def\newpage{\par\vfill\supereject} \def\newline{\hfil\break}
\def\jump{\vskip\baselineskip} \newskip\iinnffrr
\def\sjump{\iinnffrr=\baselineskip
          \divide\iinnffrr by 2 \vskip\iinnffrr}
\def\bjump{\vskip\baselineskip \vskip\baselineskip}
\newcount\nmbnota  \def\clearnmbnota{\global\nmbnota=0}
\newcount\tipbnota 

\def\note#1{\global\advance\nmbnota by 1 \ifnum\tipbnota=1
    \footnote{$^{\rm\nttlett}$}{#1} \else {\ifnum\tipbnota=2
    \footnote{$^{\nttsymb}$}{#1}
    \else\footnote{$^{\the\nmbnota}$}{#1}\fi}\fi}
\def\nttlett{\ifcase\nmbnota \or a\or b\or c\or d\or e\or f\or
g\or h\or i\or j\or k\or l\or m\or n\or o\or p\or q\or r\or
s\or t\or u\or v\or w\or y\or x\or z\fi}
\def\nttsymb{\ifcase\nmbnota \or\dag\or\sharp\or\ddag\or\star\or
\natural\or\flat\or\clubsuit\or\diamondsuit\or\heartsuit
\or\spadesuit\fi}   \clearnmbnota
\def\numberfootnote{\global\tipbnota=0} \numberfootnote
\def\setnote#1{\expandafter\xdef\csname#1\endcsname{
\ifnum\tipbnota=1 {\rm\nttlett} \else {\ifnum\tipbnota=2
{\nttsymb} \else \the\nmbnota\fi}\fi} }
 \def\endformula{\eqno\numero $$}
 \def\efr{\endformula}
\newcount\frmcount \def\clearfrmcount{\global\frmcount=0}
\def\numero{\global\advance\frmcount by 1   \ifnum\indappcount=0
  {\ifnum\cpcount <1 {\hbox{\rm (\the\frmcount )}}  \else
  {\hbox{\rm (\the\cpcount .\the\frmcount )}} \fi}  \else
  {\hbox{\rm (\applett .\the\frmcount )}} \fi}
\def\nameformula#1{\global\advance\frmcount by 1%
\ifnum\draftnum=0  {\ifnum\indappcount=0%
{\ifnum\cpcount<1\xdef\spzzttrra{(\the\frmcount )}%
\else\xdef\spzzttrra{(\the\cpcount .\the\frmcount )}\fi}%
\else\xdef\spzzttrra{(\applett .\the\frmcount )}\fi}%
\else\xdef\spzzttrra{(#1)}\fi%
\expandafter\xdef\csname#1\endcsname{\spzzttrra}
\eqno \ifnum\draftnum=0 {\ifnum\indappcount=0
  {\ifnum\cpcount <1 {\hbox{\rm (\the\frmcount )}}  \else
  {\hbox{\rm (\the\cpcount .\the\frmcount )}}\fi}   \else
  {\hbox{\rm (\applett .\the\frmcount )}} \fi} \else (#1) \fi $$}
\def\nfr{\nameformula}    \def\numali{\numero}
\def\nameali#1{\global\advance\frmcount by 1%
\ifnum\draftnum=0  {\ifnum\indappcount=0%
{\ifnum\cpcount<1\xdef\spzzttrra{(\the\frmcount )}%
\else\xdef\spzzttrra{(\the\cpcount .\the\frmcount )}\fi}%
\else\xdef\spzzttrra{(\applett .\the\frmcount )}\fi}%
\else\xdef\spzzttrra{(#1)}\fi%
\expandafter\xdef\csname#1\endcsname{\spzzttrra}
  \ifnum\draftnum=0  {\ifnum\indappcount=0
  {\ifnum\cpcount <1 {\hbox{\rm (\the\frmcount )}}  \else
  {\hbox{\rm (\the\cpcount .\the\frmcount )}}\fi}   \else
  {\hbox{\rm (\applett .\the\frmcount )}} \fi} \else (#1) \fi}
\clearfrmcount
\newcount\cpcount \def\clearcpcount{\global\cpcount=0}
\newcount\subcpcount \def\clearsubcpcount{\global\subcpcount=0}
\newcount\appcount \def\clearappcount{\global\appcount=0}
\newcount\indappcount \def\clearindappcount{\indappcount=0}
\newcount\sottoparcount 
 \newcount\draftnum \clearappcount
\clearindappcount \global\draftnum=0
\newcount\connttrre  \def\clearconnttrre{\global\connttrre=0}
\newcount\countref  \def\clearcountref{\global\countref=0}
\clearcountref
\clearsubcpcount
\clearappcount \clearindappcount
\def\references{\goodbreak\null\vbox{\jump\nobreak
   \itemitem{}{\ttaarr References} \nobreak\jump\sjump}\nobreak}
\def\beginpaper{\clearindappcount\clearappcount\clearcpcount
                  \clearsubcpcount\null}
\clearcpcount\clearcountref

%
%
\catcode`@=11
\gdef\Ref#1{\expandafter\ifx\csname @rrxx@#1\endcsname\relax%
{\global\advance\countref by 1%
\ifnum\countref>200%
\message{>>> ERROR: maximum number of references exceeded <<<}%
\expandafter\xdef\csname @rrxx@#1\endcsname{0}\else%
\expandafter\xdef\csname @rrxx@#1\endcsname{\the\countref}\fi}\fi%
\ifnum\draftnum=0 \csname @rrxx@#1\endcsname \else#1\fi}
\gdef\beginref{\ifnum\draftnum=0  \gdef\Rref{\fairef}
\gdef\endref{\scriviref} \else\relax\fi \parskip 2pt plus 2pt
\baselineskip=12pt}
\def\Reflab#1{[#1]} \gdef\Rref#1#2{\item{\Reflab{#1}}{#2}}
\gdef\endref{\relax}  \newcount\conttemp
\gdef\fairef#1#2{\expandafter\ifx\csname @rrxx@#1\endcsname\relax
{\global\conttemp=0
\message{>>> ERROR: reference [#1] not defined <<<} } \else
{\global\conttemp=\csname @rrxx@#1\endcsname } \fi
\global\advance\conttemp by 50
\global\setbox\conttemp=\hbox{#2} }
\gdef\scriviref{\clearconnttrre\conttemp=50
\loop\ifnum\connttrre<\countref \advance\conttemp by 1
\advance\connttrre by 1
\item{\Reflab{\the\connttrre}}{\unhcopy\conttemp} \repeat}
\clearcountref \clearconnttrre
\catcode`@=12
\def\slashchar#1{\setbox0=\hbox{$#1$} \dimen0=\wd0
     \setbox1=\hbox{/} \dimen1=\wd1 \ifdim\dimen0>\dimen1
      \rlap{\hbox to \dimen0{\hfil/\hfil}} #1 \else
      \rlap{\hbox to \dimen1{\hfil$#1$\hfil}} / \fi}
\ifx\oldchi\undefined \let\oldchi=\chi
  \def\cchi{{\raise 1pt\hbox{$\oldchi$}}} \let\chi=\cchi \fi
  
\def\del{\partial}   

\def\frac#1#2{{\textstyle{#1 \over #2}}}

\def\half{\ifinner {\scriptstyle {1 \over 2}}\else {1 \over 2} \fi}
  
\def\vev#1{\left\langle#1\right\rangle}

\def\simge{\rlap{\raise 2pt \hbox{$>$}}{\lower 2pt \hbox{$\sim$}}}
\def\simle{\rlap{\raise 2pt \hbox{$<$}}{\lower 2pt \hbox{$\sim$}}}
\def\buildchar#1#2#3{{\null\!\mathop{#1}\limits^{#2}_{#3}\!\null}}

\def\vbig#1#2{{\vbigd@men=#2\divide\vbigd@men by 2%
\hbox{$\left#1\vbox to \vbigd@men{}\right.\n@space$}}}

\null
%
%
%
%
\loadamsmath\chapterfont{\bfone}
\sectionfont{\scaps}\def\pb{\overline\psi}
\nopagenumbers \ifdoublepage \null\newpage \fi
{\baselineskip=12pt \line{\hfill PUPT-1308}
\line{\hfill hepth@xxx/9203028} \line{\hfill February, 1992} }
{\baselineskip=14pt \bjump\bjump
\centerline{\capsone A NOTE ON THE ZAKHAROV-SHABAT}
\sjump \centerline{\capsone TOPOLOGICAL MODEL}
\bjump\bjump \centerline{\scaps Andrea
Pasquinucci~\footnote{$^\dagger$}{Research supported by an INFN
fellowship.}}
\sjump
\centerline{\sl Joseph Henry Laboratories, Department of Physics,}
\centerline{\sl Princeton University, Princeton, NJ 08544, USA}
\vfill \centerline{\capsone ABSTRACT} \sjump
\noindent In this note I discuss some features of the topological
theory obtained from the Zakharov-Shabat
(or general $sl(2,{\Bbb C})$) hierarchy, and comment on some
possible physical and/or mathematical interpretations of it.
\bjump\bjump\bjump \eject}
\yespagenumbers\pageno=1 \beginpaper
In some recent papers (see for example refs.
[\Ref{mkdvmm},\Ref{twoarc},\Ref{moore},\Ref{noi}])
it has been shown that the double scaling
limit of hermitian and anti-hermitian matrices in the one- and two-arc
sector, leads to two dimensional quantum gravity models based on the
following $sl(2,{\Bbb C})$ hierarchies: Korteweg-de Vries (KdV),
modified KdV (mKdV), non-linear Schr\"odinger (NLS) and Zakharov-Shabat
(ZS). Of particular interest are the two dimensional quantum gravity
theories constructed from the ZS hierarchy. Indeed, the ZS hierarchy
contains both the KdV and mKdV hierarchies as reductions, and its
first (odd) critical point gives rise to a new ``topological theory"
[\Ref{moore}], different from the one obtained from the KdV hierarchy
[\Ref{grawitten},\Ref{diwitten},\Ref{topdist}]. (Notice that
only the even critical points of the ZS hierarchy transform, under
reduction, into those of the KdV hierarchy, whereas the
odd critical points don't.)

The purpose of this note is to discuss some features of the
topological theory, which will be called ``ZS topological model",
corresponding to the first critical point of the ZS
hierarchy [\Ref{moore}], and to make some comments on its
possible physical and/or mathematical interpretations. \sjump

The Zakharov-Shabat hierarchy is simply defined by introducing two
real, independent functions, $\psi$ and $\pb$, of the parameters
$(t_{-1},x,t_1,\dots)$~\note{I have adopted the conventions of ref.
[\Ref{noi}].} and by defining the flows as
$$ {\partial\psi\over\partial t_k}=\frac12(F_{k+1}-G_{k+1})
\qquad\quad {\partial\pb\over\partial t_k}=\frac12(F_{k+1}+G_{k+1})
\qquad\quad (k\geq-1) \efr
where the polynomials $F_k$ and $G_k$ are given by
$$ F_{k+1}\ =\ G'_k+(\pb-\psi)H_k\ ,\qquad \qquad
G_{k+1}\ =\ F'_k+(\pb+\psi)H_k \efr
$$  H_k'\ =\ \pb(G_k-F_k)-\psi(G_k+F_k) \efr
with
$$ F_0\ =\ \pb-\psi\ , \qquad  G_0\ =\ \psi+\pb\ , \qquad
H_0\ =\ 0\ . \efr

The connexion with the anti-hermitian 1-matrix models is given by the
following formul\ae :
$$ \vev{PP}\ = \ -\psi\pb\ ,\qquad \vev{{\cal O}_i PP}\ =\
{\del\phantom{t_i}\over\del t_i} \vev{PP} \ ,\qquad  \vev{{\cal
O}_i P}\ =\ \frac12 H_{i+1} \efr
$\vev{PP}=\del^2_x \log\, {\cal Z} = - F''$, $t_0=x$.
The string equations are
$$ \sum_{k=0}^\infty(k+1)t_k F_k \ =\ 0 \qquad
\qquad\qquad \sum_{k=0}^\infty(k+1)t_k G_k\ =\ 0\ . \efr

The string equations and the hierarchy equations can be written as
Virasoro constraints acting on the partition function of the matrix
model (or on the tau-function of the ZS hierarchy since
${\cal Z} = \tau$). These are the Virasoro constraints
of an untwisted scalar field
$$ L_n(\alpha)\, {\cal Z}\ =\ 0 \ ,\qquad\qquad\qquad n\ge -1
\nfr{virone}
where $\alpha$ is an a-priori arbitrary integration constant (see
refs. [\Ref{moore},\Ref{noi}] for a discussion of the Virasoro
constraints in the ZS hierarchy).

The topological point of the ZS hierarchy is given by
$$ t_1\ =\ \beta \qquad\qquad t_n\ =\ 0\ \ ~~~(n\neq 1) \efr
where $\beta$ is a number to be determined.
For convenience, it is better to shift $t_1$ by $-\beta$
in such a way that the topological
point is given by $t_n=0$ for any $n$.

I will also introduce the following notation for the
correlation functions:  $\vev{{\cal O}_{n_1} \dots {\cal
O}_{n_p}}_t$  will denote a correlation function computed {\sl not
at\/} the
ZS topological point (which means that it is a
function of the $t$'s),  $\vev{{\cal O}_{n_1} \dots {\cal
O}_{n_p}}$ will, instead, denote a correlation function computed {\sl
at\/} the ZS topological point (i.e. at $t_n=0$).

The Virasoro constraints, eq. \virone, can then be written
in the following way
$$\eqalignno{
\vev{P}_t = & {1\over 2\beta} \left[ \sum_{k\ge 1} (k+1) t_k
\vev{{\cal O}_{k-1}}_t + {\alpha t_0 \over 2} \right] \cr
\vev{{\cal O}_1}_t = & {1\over 2\beta} \left[ \sum_{k\ge 1} (k+1)
t_k \vev{{\cal O}_k}_t + {\alpha^2 \over 4}\right]&\nameali{virqua}\cr
\vev{{\cal O}_{n+1}}_t = & {1\over 2\beta} \left[ \sum_{k\ge 1} (k+1)
t_k \vev{{\cal O}_{k+n}}_t + \alpha \vev{{\cal O}_{n-1}}_t + \right.
\cr & \qquad \left. + \sum_{k=1}^{n-2}\left\{ \vev{{\cal O}_k}_t
\vev{{\cal O}_{n-k-2}}_t + \vev{{\cal O}_k {\cal O}_{n-k-2}}_t
\right\}\right] \ .\cr} $$

It is now necessary to consider the meaning of $\alpha$. If $\alpha$
is a parameter in the theory, from eq. \virqua\ it follows immediately
that $\alpha$ cannot assume the value zero, otherwise at the ZS
topological point all the correlation functions are zero. Moreover, to
construct a topological theory one needs to assign ghost charges to
the operators. It is not difficult to verify that a ghost charge
assignment is possible only if $\alpha=0$, leading to a trivial
theory.

The other possibility, which I will assume from now on, is that
$\alpha$ is a coupling constant, i.e. there is an operator $Q$
associated to ${\del\ \over \del \alpha}$ in the sense that
$$\vev{Q{\cal O}_{n_1}\dots {\cal O}_{n_p}}\ =\ \left.
{\del\ \over\del\alpha} \vev{{\cal O}_{n_1}\dots {\cal O}_{n_p}}_t
\right\vert_{t_n=0,\, \alpha=0}\ . \efr
Notice that if $\alpha$ is a coupling constant, then the ZS
topological  point is given by $t_n=0$ and $\alpha=0$.
As it is obvious from its definition, the $Q$ operator
has a different origin from all the other operators and understanding
its meaning could be a key to understand the ZS topological theory.

The Virasoro constraints can be recasted in a way to show explicitly
the topological recursion relations.
Consider the correlation function
$$\vev{{\cal O}_n \prod_i {\cal O}_{n_i} }\ =\ \left.
{\del\ \over\del t_n} \vev{\prod_i {\cal O}_{n_i} }_t
\right\vert_{t=0}\ .\efr
One can eliminate ${\del\ \over\del t_n}$ using the Virasoro
constraints (eqs. \virone\ or \virqua). One has:
\ifdoublepage
$$ \eqalignno{ &\vev{{\cal O}_n \prod_{i\in S} {\cal O}_{m_i} Q^p}\ =
&\nameali{rrdue}\cr &\qquad\qquad
{1\over 2\beta}  \left[  \sum_{i\in S} (m_i+1) \vev{{\cal O}_{m_i+n-1}
\prod_{j\neq i} {\cal O}_{m_j} Q^p}\ + \right.\cr & \qquad\qquad\qquad
+ \sum_{m=0}^{n-3} \vev{ {\cal O}_m {\cal O}_{n-3-m} \prod_{i\in S}
{\cal O}_{m_i} Q^p}\ + \cr &\qquad\qquad\qquad
+ \sum_{m=0}^{n-3}\,\sum_{X\cup Y=S}\, \sum_{a+b=p}
\vev{ {\cal O}_m \prod_{j\in X} {\cal O}_{m_j} Q^a} \
\cdot \cr &\qquad\qquad\qquad \qquad\qquad \cdot\
\vev{{\cal O}_{n-3-m} \prod_{k\in Y} {\cal O}_{m_k} Q^b}\
+ \cr & \qquad\qquad\qquad
\left. +\ p \vev{{\cal O}_{n-2}\prod_{i\in S} {\cal O}_{m_i}
Q^{p-1}} \right] \cr} $$
\else
$$\eqalignno{ &\vev{{\cal O}_n \prod_{i\in S} {\cal O}_{m_i} Q^p}\ =
&\nameali{rrdue}\cr &\qquad\qquad
{1\over 2\beta}  \left[  \sum_{i\in S} (m_i+1) \vev{{\cal O}_{m_i+n-1}
\prod_{j\neq i} {\cal O}_{m_j} Q^p}\ + \right.\cr & \qquad\qquad\qquad
+ \sum_{m=0}^{n-3} \vev{ {\cal O}_m {\cal O}_{n-3-m} \prod_{i\in S}
{\cal O}_{m_i} Q^p}\ + \cr &\qquad\qquad\qquad
+ \sum_{m=0}^{n-3}\,\sum_{X\cup Y=S}\, \sum_{a+b=p}
\vev{ {\cal O}_m \prod_{j\in X} {\cal O}_{m_j} Q^a} \
\vev{{\cal O}_{n-3-m} \prod_{k\in Y} {\cal O}_{m_k} Q^b}\
+ \cr & \qquad\qquad\qquad
\left. +\ p \vev{{\cal O}_{n-2}\prod_{i\in S} {\cal O}_{m_i}
Q^{p-1}} \right] \cr} $$
\fi
and the exceptions
$$ \vev{PPQ}\ =\ {1\over 4\beta}\ =\ \vev{{\cal O}_1 QQ}
\nfr{rrtre}
with the convention that ${\cal O}_{-1}\equiv 0$.

Now one has to assign ghost charges to the operators ${\cal O}_n$,
$P$ and $Q$. As usual, since $t_1$
has been shifted, one gives ghost number zero to ${\cal O}_1$.
A good assignment of ghost charges is:
$$ q({\cal O}_n)\ =\ (n-1)\qquad\qquad q(P)\ =\ -1
\qquad\qquad q(Q)\ =\ -2\ . \nfr{ghostc}

It is convenient to introduce the notation
$$ \vev{\prod_{i\in S} {\cal O}_{m_i} Q^p}_h\ ,\qquad\qquad  h\ =\
\sum_{i\in S} q({\cal O}_{m_i}) \, - 2\, p \efr
i.e. $h$ is the total ghost charge of the operators inside the
v.e.v.~. One can prove the following results:
\item{1)}{$\vev{ \cdots }_h =0 $ ~ if $h <  -4$~;}
\item{2)}{all correlation functions {\bf without} insertions of $Q$
are zero;}
\item{3)}{all correlation functions with $h\, \neq \, 4 (g-1)$
and $g=0,1,2,\dots\,$, are zero, with the exception of
$\vev{Q}_{h=-2}$~.}

\noindent Thus, there is the following ghost charge conservation
law: $$ \sum_i \, q_i\ =\ 4 (g -1) \nfr{gcons}
and I'll set $\vev{Q}=0$. I will also label the correlation functions
by the genus $g$ instead that by ghost charge $h$.

It is useful to change normalization of the operators ${\cal O}_m$ and
$Q$. Define $\widetilde{\cal O}_m$ and $\widetilde{Q}$ by:
$$ {\cal O}_m\ =\ {(m+1)!\over (2\beta)^m }\,
\widetilde{\cal O}_m \qquad m\ge 0\ ,\qquad\quad
Q\ =\ \beta \widetilde{Q}\ . \efr
Then eqs. \rrdue\ and \rrtre\ become
$$\eqalignno{& \vev{\widetilde{\cal O}_n \prod_{i\in S}
\widetilde{\cal O}_{m_i} \widetilde{Q}^p}_g\ =&\nameali{rrund}\cr
&\qquad \sum_{i\in S} {(m_i+1) (m_i+n)!\over (n+1)! (m_i+1)!}
\vev{\widetilde{\cal O}_{m_i+n-1}
\prod_{j\neq i} \widetilde{\cal O}_{m_j} \widetilde{Q}^p}_g\ + \cr
&\quad +\,(2\beta)^2 \sum_{m=0}^{n-3} {(m+1)!(n-2-m)!\over (n+1)!}
\vev{ \widetilde{\cal O}_m \widetilde{\cal O}_{n-3-m} \prod_{i\in S}
\widetilde{\cal O}_{m_i} \widetilde{Q}^p}_{g-1}\ + \cr &\quad
+\, (2\beta)^2\sum_{m=0}^{n-3}\, \sum_{X\cup Y=S}\,
\sum_{{\scriptstyle a+b=p\atop\scriptstyle g_1+g_2=g}}
 {(m+1)!(n-2-m)!\over (n+1)!}
\vev{ \widetilde{\cal O}_m \prod_{j\in X} \widetilde{\cal O}_{m_j}
\widetilde{Q}^a}_{g_1}  \cdot\cr &\quad   \qquad\qquad \cdot\
\vev{\widetilde{\cal O}_{n-3-m} \prod_{k\in Y}
\widetilde{\cal O}_{m_k} \widetilde{Q}^b}_{g_2}\
+ \cr &\quad +\ 2p\,{(n-1)!\over (n+1)!}
\vev{\widetilde{\cal O}_{n-2}\prod_{i\in S}
\widetilde{\cal O}_{m_i} \widetilde{Q}^{p-1}}_g \cr} $$
and
$$ \vev{PP\widetilde{Q}}_{g=0}\ =\ {1\over 4\beta^2}\ =
\ \vev{\widetilde{\cal O}_1 \widetilde{Q}\widetilde{Q}}_{g=0}\ .
\nfr{rrdod}
In particular
$$\eqalignno{
& \vev{P \prod_{i\in S} \widetilde{\cal O}_{m_i} \widetilde{Q}^p}_g
\ =\ \sum_{i\in S} \vev{\widetilde{\cal O}_{m_i-1}
\prod_{j\neq i} \widetilde{\cal O}_{m_j} \widetilde{Q}^p}_g
&\nameali{rrtred}\cr & \vev{\widetilde{\cal O}_1 \prod_{i=1}^{s-p}
\widetilde{\cal O}_{m_i} \widetilde{Q}^p}_g\ = \left(2g -2 +s\right)
\vev{\prod_{i=1}^{s-p} \widetilde{\cal O}_{m_i} \widetilde{Q}^p}_g
&\nameali{rrquad}\cr} $$
where, in the last formula, the ghost charge conservation
has been used.\sjump

{}From eq. \rrtred\ it follows that $P$ could play a r\^ole similar to
that of the {\sl puncture}\/ operator in the standard (KdV)
topological gravity [\Ref{grawitten},\Ref{diwitten},\Ref{topdist}],
with the $\widetilde{\cal O}_m$ ($m\geq 1$)
acting as ``descendants" and $\widetilde{Q}$ as a kind of second
``primary" operator.
Moreover, eq. \rrquad\ tells us the $\widetilde{\cal O}_1$ behaves
exactly as the {\sl dilaton\/} operator in the standard (KdV)
topological gravity, with the only exception of the correlation
function $\vev{\widetilde{\cal O}_1
\widetilde{Q}\widetilde{Q}}_{g=0}$.
Anyway, it is clear from the recursion relations eq. \rrund\ that the
ZS topological gravity has very different characteristics from the
usual KdV topological gravity due to the distinctive r\^ole played by
the operator $Q$.

Let first try to make a comparison between the ZS topological gravity
and the well known KdV topological gravity [\Ref{diwitten}].
Behind the $Q$
operator, also ${\cal O}_1$ shows a different behaviour.
Indeed, in KdV topological gravity
${\cal O}_1$ is a descendant operator, which means that every
correlation function in which it appears can be turn, using the
recursion relations, into a correlation function involving only
primary operators. This is not what happens in eq. \rrdod. Thus,
it is not clear if the standard definition of primary and
descendant operators can be applied to the ZS topological model.

Indeed, one could think of $Q$, $P$ and $R\
\buildchar{=}{\rm def}{ }\ {\cal O}_1$ as ``primary" operators.
In this case, on the Small-Phase-Space at genus zero, one has
$$ \log\,{\cal Z}\ =\ -F\ =\ {t_Q (t_P)^2 \over 8(\beta -t_R)} +
{(t_q)^2 \over 8\beta}\,\log\left(1-{t_R\over \beta}\right) +
\frac12 \delta (t_Q)^2 \efr
where $\delta$ is an arbitrary integration constant which must be set
to zero if $\vev{QQ}=0$, and the constitutive equations:
$$ \eqalignno{ \vev{PP}_t\ &=\ {t_Q \over 4(\beta-t_R)}\cr
\vev{PQ}_t\ &=\ {t_P\over 4(\beta-t_R)} \cr
\vev{QQ}_t\ &=\ -\frac14 \log \left( 1- {t_R\over \beta}\right)
+\,\delta &\nameali{treuno}\cr
\vev{RR}_t\ &=\ 16\vev{PP}_t\left(\vev{PQ}_t\right)^2 + 2
\left(\vev{PP}_t\right)^2 \cr
\vev{PR}_t\ &=\ 4\vev{PP}_t\vev{PQ}_t \cr
\vev{RQ}_t\ &=\ 2\left(\vev{PQ}_t\right)^2 +\,\vev{PP}_t\ . \cr}
$$ If $\vev{PP}_t$ would had been exchanged with $\vev{RQ}_t$, these
constitutive equations would had resembled those for a three-primary
KdV topological model.

To understand the behaviour of the $Q$ and ${\cal O}_1$ operators,
another possibility [\Ref{comwitt}] is to consider the
${\cal O}_m$ ($m\geq 0$)
operators as linear combinations of operators coming from the
topological gravitational sector and a topological matter sector,
for example a topological $U(1)$-gauge theory. Although this
interpretation is very
promising and the behaviour of ${\cal O}_1$ can be easily understood,
I haven't yet found a consistent way of applying it.

Another possible way of understanding the ZS topological model is that
of looking for a mathematical description~\note{I thank E.~Witten for
his patience in explaining to me many mathematical details on the
algebro-geometrical formulation of topological gravity.} in terms of
an intersection theory on Riemann surfaces
[\Ref{topwitten},\Ref{mathwitten},\Ref{robtop}].
An intersection theory on Riemann surfaces which could be a good
candidate for being the algebro-geometrical description of the ZS
topological model can be constructed as follows [\Ref{comwitt}].

Consider a Riemann surface $\Sigma$
with $g$ handles  and $n=s+t$ marked points, and its Deligne-Mumford
compactified moduli space
$\overline{{\bfcal M}}_{g,n}$. By the Riemann-Roch theorem this is a
space of complex dimension
$ {\rm dim}\, \overline{{\bfcal M}}_{g,n} = 3g-3+n $.
Consider the line bundle
$$ {\bfmath L}\ \buildchar{=}{\rm def}{ }\
\buildchar{\scriptstyle \bigotimes}{\scriptstyle s}{\scriptstyle i=1}
\, {\bfcal O}(q_i)^{-1} \,
\buildchar{\scriptstyle \bigotimes}{\scriptstyle t}{\scriptstyle j=1}
\, {\bfcal O}(p_j)^{n_j}\ , \qquad n_j\geq 0 \efr
over $\overline{{\bfcal M}}_{g,n}$. That is, its sections are
functions that can have poles of order $n_j$
at the point $p_j$ or zero of first order at the point $q_i$. The
degree of ${\bfmath L}$ is
$ {\rm deg}\, ({\bfmath L}) = \sum n_j -s $.
Consider now a $D$-dimensional vector bundle $V$ over the moduli
space, whose fiber $V_\Sigma$ at a point $\Sigma\in
\overline{{\bfcal M}}_{g,n}$ is
defined as the space of holomorphic sections of ${\bfmath L}$~:
$$ V_\Sigma\ \buildchar{=}{\rm def}{ }\ H^0(\Sigma, {\bfmath L})
\qquad\quad {\rm if}\ \ H^1(\Sigma, {\bfmath L})\ =\ 0\ . \efr
(This definition is correct only if $H^1(\Sigma, {\bfmath L}) = 0$.
A more general definition that takes into account that
$H^1(\Sigma, {\bfmath L})$ may not always vanish, can be done
following ref. [\Ref{mathwitten}].) This
vector bundle has Chern classes and in particular the Top Chern class
$c_{top}(V)= c_{D}(V)$. Notice that the Riemann-Roch theorem tells us
that the fiber of $V$ has complex dimension $D = 1-g + \sum n_j -s$.
Now, the possible connexion with the ZS topological model is given by
the following formula:
$$ \vev{\prod_{j=i}^t {\cal O}_{n_j} \cdot\, Q^s}_g\
\buildchar{=}{?}{ } \  \vev{c_D(V)\, ,\,
\overline{{\bfcal M}}_{g,n} }\ . \efr
To investigate if this conjecture is correct, one must study the
properties of the intersection numbers $\vev{c_D(V)\, ,\,
\overline{{\bfcal M}}_{g,n} }$
and compare them to those of the ZS topological correlation functions.
These intersection numbers vanish unless a dimensional condition is
obeyed, namely $\sum_i \, q_i = 4 (g -1)$.
This is exactly the ghost selection rule \gcons . A few examples of
intersection numbers permitted by this rule are:
$$ \eqalignno{D\ =\ 0\qquad\qquad\qquad & \vev{{\cal O}_1 Q^2}_0
\ =\ 1\ =\ \vev{P^2 Q}_0 &\numali\cr
D\ =\ 1\qquad\qquad\qquad & \vev{{\cal O}_3 Q^3}_0\ ,
\ \ \ \ \vev{{\cal O}_1^2 Q^2}_0\
,\ \ \ \ \vev{P{\cal O}_2Q^2}_0 \cr
& \vev{P^2{\cal O}_1Q}_0\ ,\ \ \ \ \ \ \vev{P^4}_0\ ,\ \ \
\ \vev{{\cal O}_1}_1\cr
\dots \qquad\qquad\qquad\quad\  &\cr}$$
where I have actually written the correlation functions corresponding
to the intersection numbers (i.e. $\vev{{\cal O}_1Q^2}_0$ stays for
$\vev{c_0(V)\, , \, \overline{{\bfcal M}}_{0,3} }$ where ${\bfmath L}
= {\bfcal O}(q_1)^{-1}\otimes {\bfcal O}(q_2)^{-1}\otimes
{\bfcal O}(p_1)^{+1} $, etc.).

Although the situation up to now looks promising, many thing must be
proven in the mathematical theory even before proposing this
intersection theory as the algebro-geometrical description of
the ZS topological model. For example, one should prove that
all the intersection numbers without insertions of $Q$
(i.e. with $s=0$) are zero and that at least eqs. \rrtred\  and
\rrquad\ hold. To do that, one has to understand the
meaning of ``descendant" operator from the mathematical point of view,
and this requires the full mathematical apparatus of ref.
[\Ref{mathwitten}] (especially section 2).

In conclusion, in this note I have discussed some properties and
possible interpretations of the ZS topological theory. This theory,
although similar in many aspects to the (pure) KdV topological
gravity, has various interesting new features and its physical and/or
mathematical interpretation is still unclear. Obviously, it will be
very interesting to get a better understanding of this new two
dimensional topological theory.\jump

I would like to thank Edward Witten for stimulating discussions
and suggestions, Cesare Reina for a mathematical explanation,
Chiara Nappi for her constant encouragement and
Tim Hollowood for his collaboration at the early stages of this work.
\references
\beginref
\Rref{noi}{T.~Hollowood, L.~Miramontes, A.~Pasquinucci and C.~Nappi,
``{\sl Hermitian versus anti-hermitian 1-matrix models and their
hierarchies\/}", preprint IASSNS-HEP-91/59, PUPT-1280, September 1991,
to be published in Nucl. Phys. {\bf B}.}
\Rref{moore}{\v C.~Crnkovi\'c, M.~Douglas and G.~Moore, ``{\sl Loop
equations and the topological phase of multi-cut matrix models\/}",
preprint YCTP-P25-91, RU-91-36, August 1991.}
\Rref{grawitten}{E. Witten, ``{\sl On the Structure of the Topological
Phase of two dimensional Gravity\/}", Nucl. Phys. {\bf B340} (1990)
281.}
\Rref{diwitten}{R.~Dijkgraff and E.~Witten,
``{\sl Mean field theory, topological
field theory and multi-matrix models\/}", Nucl. Phys. {\bf B342}
(1990) 486.}
\Rref{topdist}{J.~Distler,
``{\sl 2d quantum gravity, topological field theory and
multi-critical matrix models\/}", Nucl. Phys. {\bf B342} (1990) 523.}
\Rref{mkdvmm}{\v C. Crnkovi\'c and G. Moore, ``{\sl Multi-critical
multi-cut matrix models\/}", Phys. Lett. {\bf257B} (1991) 322.}
\Rref{twoarc}{P.~Mathieu and D.~S\'en\'echal,
``{\sl A well-defined multi-critical
series in hermitian one matrix models\/}", Phys. Lett. {\bf 267B}
(1991) 475.}
\Rref{topwitten}{E.~Witten, ``{\sl Two dimensional gravity and
intersection theory on moduli space\/}", Surveys in Differential
Geometry {\bf 1} (1991) 243.}
\Rref{mathwitten}{E.~Witten, ``{\sl Algebraic geometry associated
with matrix models of two dimensional gravity\/}", preprint
IASSNS-HEP-91/74, October 1991.}
\Rref{robtop}{R.~Dijkgraaf, ``{\sl Intersection theory, integrable
hierarchies and topological field theory\/}", preprint
IASSNS-HEP-91/91, December 1991.}
\Rref{comwitt}{E.~Witten, private communication.}
\endref
\ciao
